\documentclass[12pt,preprint]{aastex}
\usepackage{natbib}
\usepackage{amsmath}
\shorttitle{The Black Hole Masses of Radio Quasars}
\shortauthors{Oshlack, Webster, Whiting}

\begin{document}

\title{Black Hole Mass Estimates of Radio Selected Quasars}

\author{A. Y. K. N. Oshlack,}
\affil{School of Physics, University of Melbourne, Parkville, Victoria 3010, Australia}
\email{aoshlack@physics.unimelb.edu.au}

\author{R. L. Webster}
\affil{School of Physics, University of Melbourne, Victoria 3010, Australia}
\email{rwebster@physics.unimelb.edu.au}
 
\author{and M. T. Whiting}
\affil{School of Physics, University of Melbourne, Victoria 3010, Australia}
\email{mwhiting@physics.unimelb.edu.au}

\begin{abstract}

The black hole (BH) mass in the centre of AGN has been estimated for a
sample of radio-selected flat-spectrum quasars to investigate the relationship
between BH mass and radio properties of quasars. We have used the
virial assumption with measurements of the H$\beta$ FWHM and luminosity
to estimate the central BH mass. In contrast to previous
studies we find no correlation
between BH mass and radio power in these AGN. We find a range in BH
mass similar to that seen in radio-quiet quasars from previous
studies. We believe the reason that the low BH mass radio-loud quasars
have not been measured in previous studies is due to optical selection
effects which tend to miss the less optically luminous radio-loud sources.

\end{abstract}


\section{Introduction}

Although it is commonly accepted that quasars harbor a super-massive
black hole (BH) in their core, neither the
distribution of the physical
characteristics of these BHs, nor their relationship to other
properties of quasars is clear. In this paper we further explore the
connection between BH mass and radio luminosity in
a quasar. 

Recently a range of methods have been used to
determine the BH masses in galaxies and quasars. For nearby galaxies,
stellar velocity dispersions and high resolution optical images 
are used to determine the BH  and bulge masses. 
Recently a tight correlation between BH mass
and velocity dispersion in the bulge of galaxies has been found
\citep[$M_{BH} - \sigma$ relation:][]{geb00,fer00}. In addition there
is evidence 
that the mass of the bulge component of the host galaxy
correlates with BH mass \citep{kor95,mag98}.
In AGN, it is difficult to determine BH mass using host galaxy dynamics
or structure as the nucleus swamps the light from the host galaxy. Also
since the redshifts are large, the projected sizes of the host
galaxies are relatively small. Therefore indirect
methods have been developed to
measure BH mass in quasars.

The mass of the BH in a quasar can be estimated using the virial technique.
\citet{lao98} assumed that the H$\beta$-emitting clouds 
in the broad line region (BLR) are virialized, and then required
only estimates of the radius of the H$\beta$-emitting region and
the rotational velocity of that region to determine the mass.
In quasars, the radius of the BLR has also been 
estimated using reverberation
techniques. This method tracks intensity variations in the continuum
luminosity which, a certain time later, are reflected in the emission line
flux. This time lag is interpreted as the light travel time between
the core of the quasar and the broad emission line
region and therefore gives the radius of the broad line clouds from
the central continuum source. 
Reverberation mapping measurements have been restricted to a
few dozen Seyferts and quasars as they require intensive regular
monitoring over periods of years and highly variable
quasars. The quasars so far studied may
not represent the broader AGN population. Regardless of this, these
observations give  extremely
interesting results. The time delay for
different ionization species provides strong support for photoionization
of the broad line region and evidence that the kinematics of the broad
line region is dominated by the central BH \citep{pet00a}.

The BLR clouds are assumed to have velocities related
to the widths of the broad emission lines so if the velocities of the
clouds are assumed to be Keplerian, the mass of the central BH can be
deduced. 
Several low luminosity quasars have been studied using both
reverberation techniques and galactic velocity dispersion to measure
their BH mass \citep{fer01}. These results show that
both these methods give consistent BH masses and that the
M$-\sigma$ relation holds for both active and non-active galaxies.

Estimates of the BH masses of radio-loud quasars have been published for
several samples. \citep{lao00} used the Palomar-Green sample of quasars
\citep[PG;][]{sch83} with spectra provided by \citet{bor92} to deduce
that a relatively powerful 
jet requires a BH with mass $M_{BH} > 3 \times 10^8 $M$_{\odot}$.
The key point Laor makes is that the BH mass does not depend on radio
luminosity, but on the radio luminosity {\sl relative} to the optical
luminosity (denoted by the ratio $\cal R$).
The PG sample comprises quasars selected for their UV-excess, and
a limiting magnitude $B \lesssim 16.6$ giving a population
biased towards low redshift quasars.

\citet{gu01} consider a sample of very radio-loud quasars taken from
the 1Jy, S4 and S5 catalogues. They estimate the BH masses for these quasars
also using the virial technique. They are able to divide their sample
into steep spectrum and flat spectrum sources and find that the
distribution of properties is similar for both subsets. They find 7
out of their 86 sources have BH masses $<10^8$M$_{\odot}$ unlike
previous authors \citep[eg,][]{lao00}.
In another carefully matched sample of radio-galaxies, radio-loud and
radio-quiet quasars, \citet{dun01} show that the host galaxies of each
class of AGN are nearly indistinguishable. However these authors
confirm the \citet{lao00} result that radio-loud quasars have BH masses $
\gtrsim 10^9$M$_{\odot}$.

A second line of enquiry relates radio continuum emission to the mass
of the BH. \citet{lac01} used the First Bright
Quasar Survey \citep[FBQS;][]{whi00} in their analysis.  These quasars
are the optically brightest ($R\leq 17.8$), blue ($B-R \leq 2.0$)
quasars in a radio sample selected at $\lambda = 20cm$.  These authors obtained a best fit relation
\begin{equation}
log L_{5GHz} = 1.9 Log M_{BH} + 1.0 log(\frac{L}{L_{Edd}}) +7.9
\end{equation}
giving a continuous, monotonic, dependence of radio luminosity on BH mass.
These studies used the virial estimator for BH mass.
This is similar to the result by \citet{fra98} who found a very steep
dependence of BH mass on radio flux, $P_{5GHz}
\propto M_{BH}^{2.5}$, and attributed this to BH accretion in an
advection dominated accretion flow.

Most recently \citet{ho01} has compiled a complete list of BH
measurements which use robust methods to determine mass (stellar
dynamics, reverberation mapping or maser dynamics). He finds that the
BH mass shows little dependence on $\cal R$ and that the radio luminosity
correlates poorly with BH mass.

We have been studying the properties of the Parkes Half-Jansky 
flat-spectrum Sample \citep[PHFS;][]{dri97}, to better understand 
the emission mechanisms 
particularly at optical wavelengths.  We find that the optical
emission in radio quasars is a mixture of several different
components: a blue component similar to the `big blue bump'
in optically-selected quasars, which is presumed to photoionize
the BLR, plus a synchrotron component which turns over somewhere
in the optical-to-IR and is a significant
contributor to the optical emission in a large fraction of sources
\citep{whi01}. Two major
differences between our sample and those already published are
(1) our sample is completely identified in the optical, and
we have not selected optically bright sources, and (2) our sources
are flat spectrum, which suggests that the radio flux may be
be boosted by orientation effects.

In section~\ref{meth}, we discuss the method used for estimation of BH
mass from 
measured optical parameters. In section~\ref{data}, we describe our 
application of these techniques to the PHFS dataset. Results and discussion
are given in section~\ref{disc}.  Where required, a cosmology with $H_0
=75 $ km s$^{-1}$ Mpc$^{-1}$ and q$_0=0.5$ is assumed.

\section{Method}
\label{meth}

When the motions of the gas moving around the BH are dominated by the
gravitational forces,
virial estimates of the BH mass can be made based on the radius and
velocity of this gas using $M_{BH}=rv^{2}G^{-1}$. 
There is evidence that the BLR gas emitting the H$\beta$ line is
virialized \citep{pet00a}. To estimate BH masses for this sample, we follow
the work of \citet{kas00} who used
reverberation mapping to derive an empirical relationship between radius and
luminosity.
To determine $v$, the rotational velocity, we correct
full-width half-maximum $v_{FWHM}$ of the H$\beta$ emission line by a factor of
$\sqrt{3}/2$ to account for random orientation affecting the width of
emission lines. The assumption of random orientation is discussed
further in Section \ref{bias}. The mass is then
\begin{equation}
M = 1.464 \times 10^{5} \left(\frac{R_{BLR}}{\text{
lt-days}}\right)\left(\frac{v_{FWHM}}{10^{3}{\rm km~s}^{-1}}\right)^{2}
M_{\odot}
\end{equation} 

The empirical relation of \citet{kas00}
relates the radius
of the emission lines to the continuum luminosity of the
source:
\begin{equation}
R_{BLR} = (32.9)\left[\frac{\lambda L_{\lambda}(5100
{\text \AA})}{10^{44} {\text erg~s}^{-1}}\right]^{0.700} \text{lt-days}
\end{equation}
with an error in the fitted slope of this relation of $<5\%$. This
correlation holds above a luminosity of $10^{44}$ erg s$^{-1}$. Below
this value the correlation is less certain and at low luminosity most
values of the emission line region radius lie well below the linear relation,
which would lead to an even lower estimate of the BH mass.
 
Although the slope of this relation cannot be explained theoretically,
the radius of the BLR is approximately linear with
the strength of the radiation field ionizing the BLR gas.
Therefore, to use the H$\beta$ BH mass estimator, two measurements are
involved. Firstly, the luminosity at 5100\AA\ and secondly, the
H$\beta$ FWHM. 

\section{PHFS Data}
\label{data}

The data used in the analysis are taken from the PHFS \citep{dri97}. 
The selection criteria for the sample are:
(1) Radio loud: 2.7GHz flux $>$ 0.5 Jy;
(2) Flat spectrum taken from the 2.7~GHz and 5~GHz fluxes: $\alpha >
-0.5$, where $S_\nu \propto \nu^\alpha$;
(3) Dec range +10 to $-$45 and galactic latitude $|$b$| > 10^{\circ}$.
The selection criteria produced 323 radio sources where optical
identification has been made for 321 sources. Therefore
the sample has no optical selection criteria.
The flat spectrum criterion may mean that sources where
the radio jets are predominantly
oriented towards us are preferentially selected.
Most of the radio morphologies are compact. 

For the PHFS, estimating the optical luminosity of the quasar is difficult as
there is a very large range in colour for the sources \citep{fra00}. This
means we cannot use a general K-correction for all the objects. In the
initial plots  we will use the B$_j$ magnitude to calculate the
luminosity. The error in the B$_j$ magnitudes from the COSMOS catalogue
is quoted  as being $\pm 0.5$ magnitudes, but \citet{dri97} have found
that  some
fields seem to be in error by more than one magnitude. This produces
an error in the BH mass estimates of about a factor of 2.
 Given that the range of masses covers 4 orders of
magnitude, these errors are not critical. In section \ref{bias} we only
consider the subset of quasars which have more accurate magnitudes. 
The radio flux of the sources given in \citet{dri97}
comes from the Parkes catalogue (PKSCAT90, \citet{wri90}) at 5GHz.
It should also be noted that there is a large range in B$_j$
magnitudes, including very faint sources which have luminosities in the Seyfert
regime.

187 sources in the sample have
low to moderate resolution
spectra (discussed in \citet{fra01}). 
Of these 101 were obtained at the AAT and Siding Spring 2.3m
telescopes \citep{dri97}, and 86 are from the compilation of \citet{wil83}. The
quality and wavelength coverage of the spectra is diverse; many have
low resolution and signal-to-noise. Estimates of the likely errors
are further discussed in the next section.

\begin{figure}
\includegraphics[angle=-90]{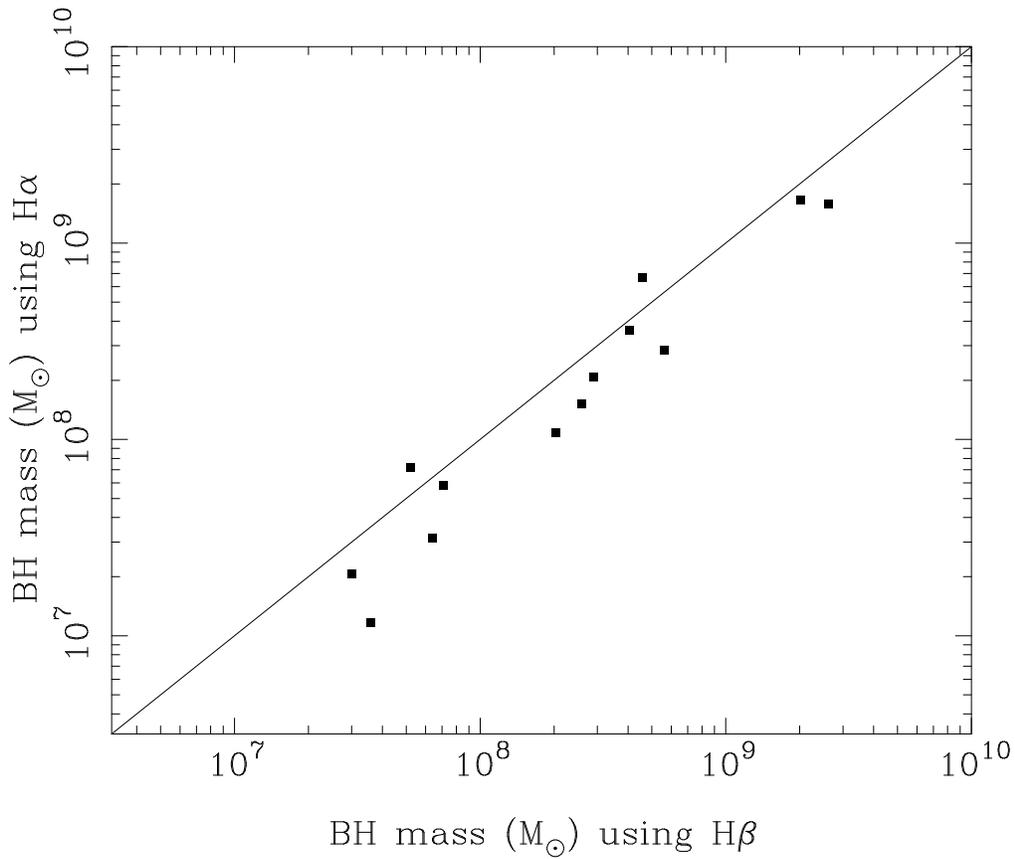}
\caption{ A comparison of BH masses estimated from using H$\alpha$
  and H$\beta$ velocity widths for the same quasars. It can be seen
  that the BH estimates using H$\alpha$ are consistently lower by an
  average of 0.157 dex. The line represents equal BH mass.}
\label{HaHb}
\end{figure}

All spectra with a wavelength range covering H$\beta$ were used in the
analysis. This gave a list of 39 quasars. Each  H$\beta$ emission
line was fitted by a Gaussian and Lorentzian using the {\sl splot}
package in IRAF. If H$\beta$ was outside the observed wavelength range
but H$\alpha$ was observed then the
fit was done for
H$\alpha$. The Lorentzian fits were better
approximations to the line shapes by eye. The Gaussian fits consistently
underestimated the peak flux; therefore Lorentzian fits had consistently
lower FWHMs. In the analysis we used the
Lorentzian fits for measurements of the line widths.
The errors in the estimation of $v_{FWHM}$ are of the order of
10-20\%. Where there was both the H$\beta$ and H$\alpha$ lines
measured we calculated the BH mass using both values for the FWHM. The
results of this is shown in Figure~\ref{HaHb}. It can be seen that the
H$\alpha$  widths give consistently lower BH masses, on average,
0.157~dex lower. 
Therefore a correction of $10^{0.157}$ was made to
all BH masses estimated using H$\alpha$.

The radio loudness parameter $\cal R$ is defined by ${f_{\nu}(5GHz)}/
f_{\nu}$(5100\AA).
A value of ${\cal R} = 10$  is usually taken as the break in a
bimodal distribution, dividing radio-loud and radio-quiet.
All the PHFS sources are clearly radio-loud with $10^2 < {\cal R} < 10^5$.
Table~1 gives the observed parameters for each source.  

\section{Results and Discussion}
\label{disc}

Considerable care must be exercised in searching for correlations between
observed and derived quantities, particularly when the plotted quantities
both depend in a similar way on distance.  As an example, in a radio
flux-limited sample, most quasars have a flux near the sample limit.
Therefore two quantities which depend on distance will be correlated,
including BH mass calculated from the virial estimator described
above, which uses the absolute luminosity.

\subsection{Correlations with Black Hole Mass}

\begin{figure}
\includegraphics[angle=-90]{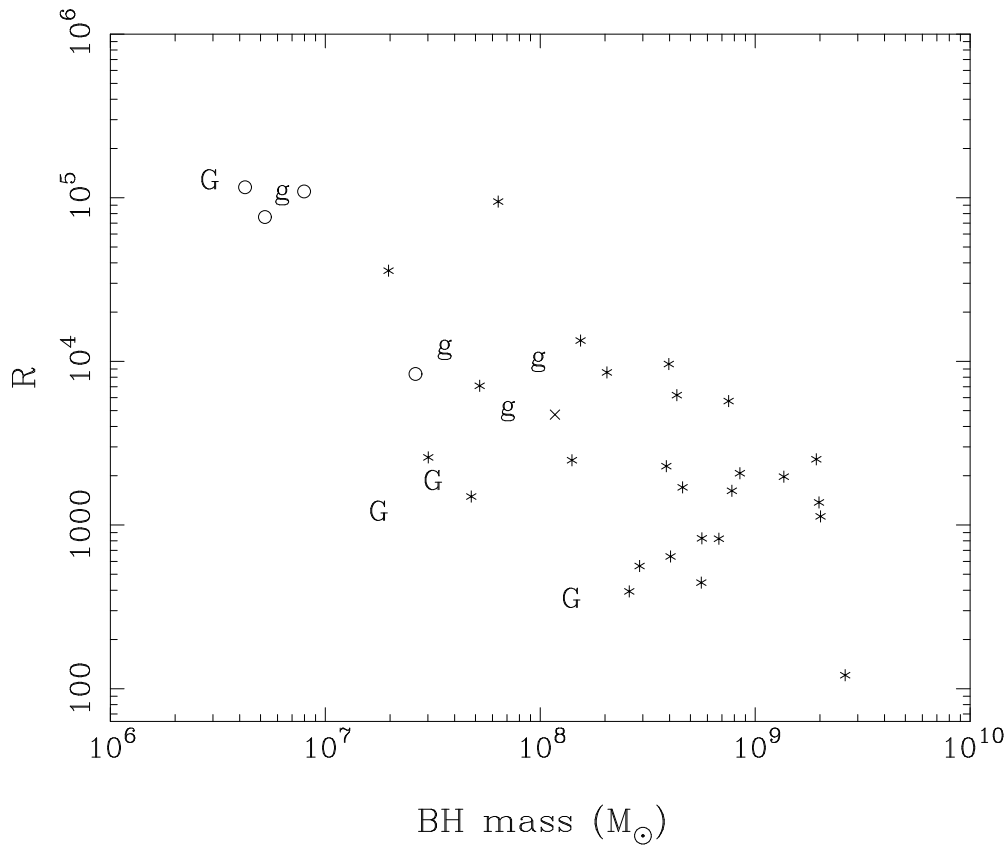}
\caption{The BH mass vs $\cal R$ (the ratio of radio to optical
flux).  This plot shows a large range in BH mass. The
stars represent BH mass calculated using H$\beta$ FWHM while the
circles are calculated using H$\alpha$ and the cross uses H$\gamma$. 8
of the objects in the sample have been classified as galaxies or
merger events on the optical plates meaning that the host galaxy is
visible. These objects have been represented with a `g' when using
H$\beta$ to calculate the BH mass and a `G' when using
H$\alpha$. These objects also span a range at the lower end of BH mass.}
\label{bh_R}
\end{figure}

Figure~\ref{bh_R} plots 
$\cal R$, the ratio of radio to optical flux, against the BH mass of
each quasar. 
Contrary to some previous results but consistent with \citet{ho01}, we
find a large range in derived BH masses. The apparent anti-correlation
here is a consequence of using the 
optical flux in the measurement of BH mass and also using it in the
calculation of $\cal R$. A significant fraction of the
masses are well below limit of $3\times 10^{8}
M_{\odot}$ suggested by \citet{lao00} as a necessary condition for
radio-loud quasars. In fact, the range in BH masses is similar
to the radio-quiet population of the PG quasar sample \citep{lao00}.
For 8 of the AGN the host galaxies are visible but it is
clear from figure~\ref{bh_R}, where these galaxies are marked with a
`g' or `G', that these are not the only low
BH mass. In these cases we expect the quasar luminosity to be
over-estimated, due to the inclusion of galaxy flux. The likely
increase in flux is $<50\%$ \citep{masci98}. Only
two of the sources have redshifts $<0.1$.

\begin{figure}
\includegraphics[angle=-90]{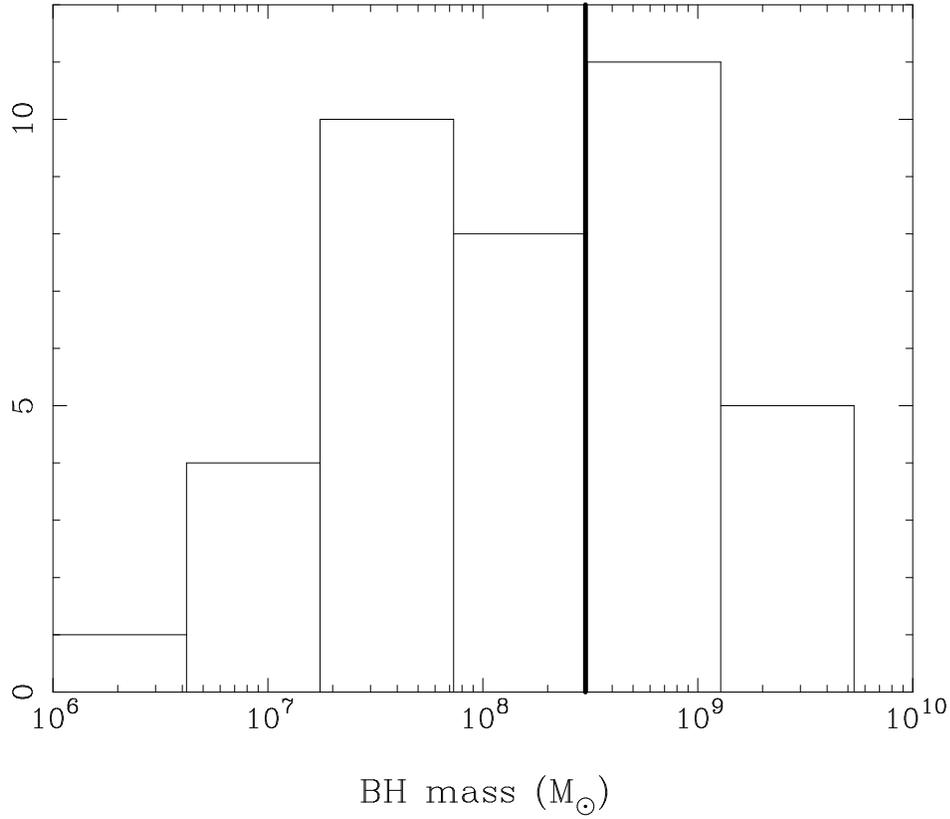}
\caption{A histogram of the estimated BH mass for radio loud quasars
from the PHFS. The solid line represents the lower limit to BH mass
stated by \citet{lao00} as necessary for having radio-loud quasars.}
\label{bhmass}
\end{figure}

\begin{figure}
\includegraphics[angle=-90]{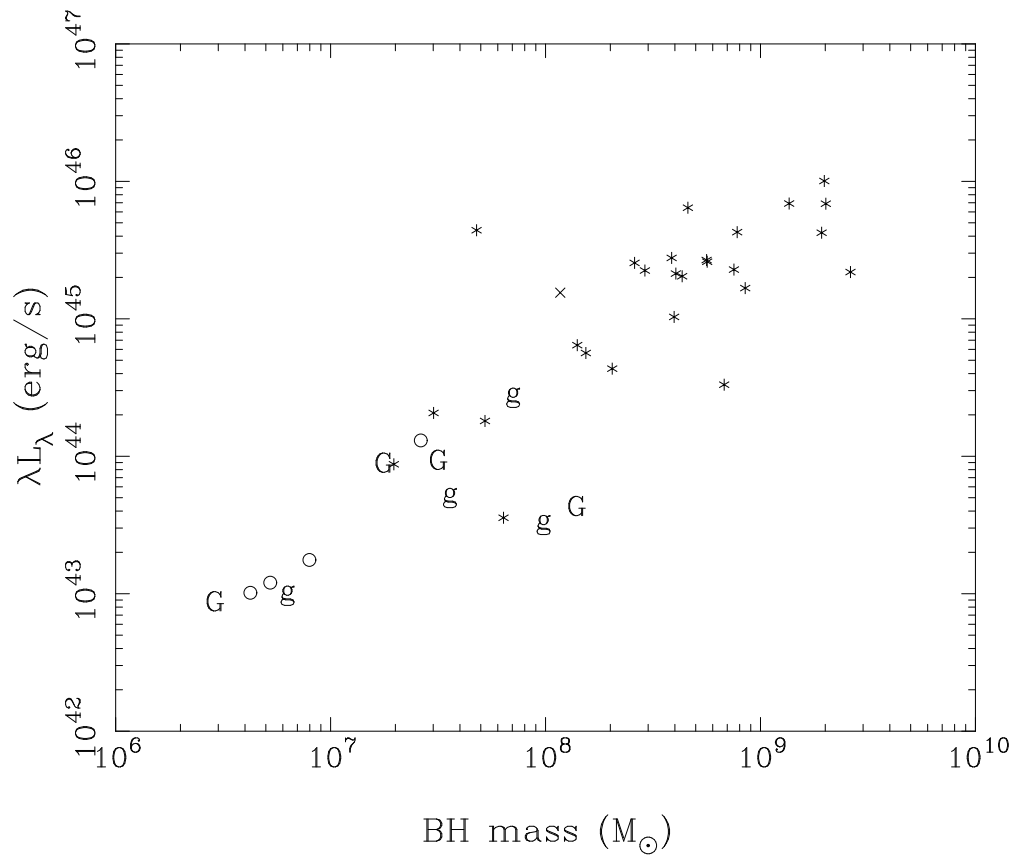}
\caption{The correlation between BH mass and optical luminosity
demonstrates the dependence of the virial estimate on the
luminosity. It shows that the low BH mass quasars are a consequence of
using the optical luminosity to estimate the radius of the emission
line gas and the velocity width of the emission line only leads to
a small scatter in the relation even though it has a higher order
dependence in the virial assumption. The symbols as for Fig~\ref{bh_R}.}
\label{bh_opt}
\end{figure}

A histogram of the BH mass is shown in Figure~\ref{bhmass}.
The low values for the BH mass are a direct consequence of the
low optical luminosities, with the range in the width of the
H$\beta$ emission line introducing some scatter.  This dependence
is clearly shown in Figure~\ref{bh_opt} where the optical luminosity is plotted
against the BH mass, and there is no evidence of
a lower limit in either optical luminosity or BH mass for any of these
radio-loud quasars. 

\begin{figure}
\includegraphics[angle=-90]{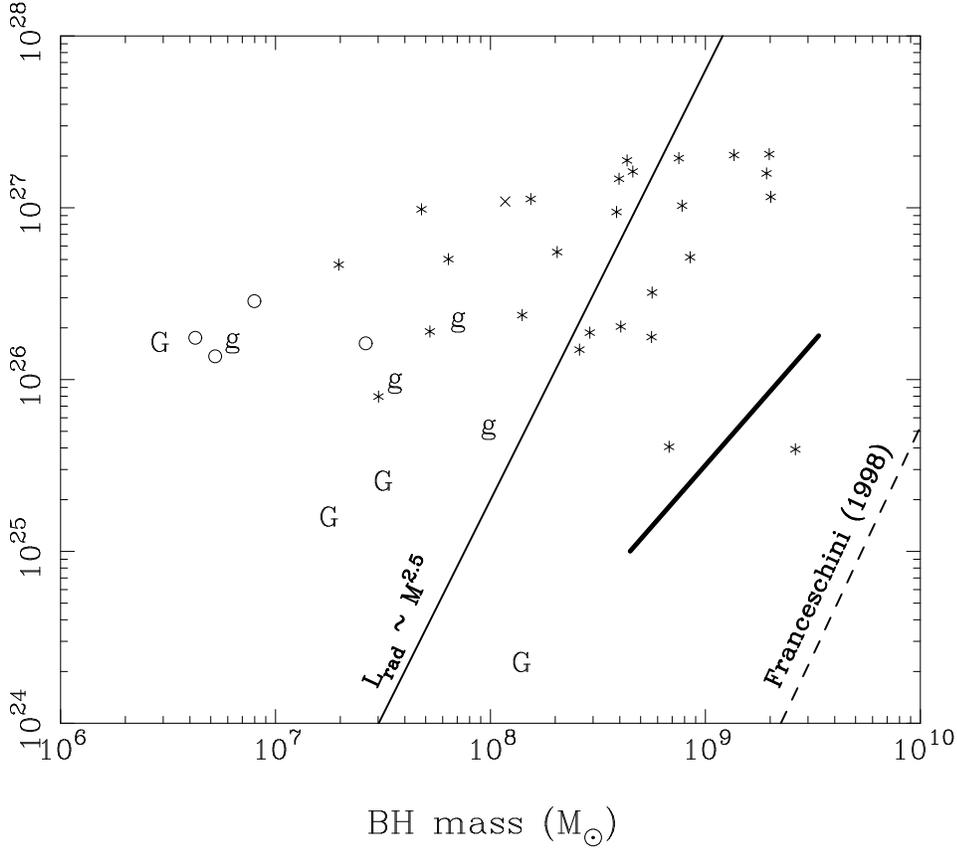}
\caption{A plot of BH mass vs Radio power at 5 GHz. There is no evidence for a
  correlation or lower limit to the distribution. All the sources are
  powerful radio sources. The thick solid line represents constant, observed
  flux over different redshifts. The dashed line represents the
  correlation found by \citet{fra98} and the thin solid line
  corresponds to the radio power going like the BH
  mass to the power of 2.5 ($L_{rad} \sim M_{\odot}^{2.5}$) similar to
  the upper boundary suggested by \citet{dun01}. Symbols as for
  Figure~\ref{bh_R}.} 
\label{bh_rad}
\end{figure}

Figure~\ref{bh_rad} shows the relationship between radio power and 
BH~mass. There is no clear relationship between these
variables unlike the results of \citet{fra98} and \citet{lac01}. The
fact that we
are not seeing a relationship between BH mass and radio power is a
consequence of the fact that there isn't a tight
correlation between optical luminosity and radio luminosity in
quasars \citep{ste00,wad99}. This has been investigated by
\citet{hoo96} for the LBQS who
find that the distribution of radio luminosity does not depend on
absolute magnitude over most of the range of M$_B$. Furthermore
the PG survey used by \citet{lao00} differs from this and other
optically selected samples in its population of radio-loud quasars
with respect to optical luminosity \citep{hoo96}. The PG sample
contains a higher fraction of radio-loud quasars which have bright
absolute magnitudes compared to other surveys like the LBQS
\citep{hoo96}. If we consider
Figure 2 in \citet{ho01}, then our data points lie in the
underpopulated upper left region of this plot. This confirms
Ho's conclusion that there is no obvious relationship
between these two variables. However we note that our sources are flat
spectrum and the radio-flux may be boosted by beaming which would tend
to increase the value of $\cal R$. Also included
in Figure~\ref{bh_rad} is a line of constant observed flux. This is
included to demonstrate the correlation that would be observed for objects with
the same measured flux (eg: the flux limit of the sample where the
majority of object will be selected)
at the different redshifts of the sources.

\subsection{Possible Biases in Estimates of Parameters}
\label{bias}

There are two measurements involved in deriving the BH mass for any
object while using this method: firstly the optical flux, to derive the
radius to the BLR and secondly the velocity width, $v_{FWHM}$ used for
the velocity component in the virial assumption. There are several
factors which might affect the applicability of the virial method to the
quasars we are considering.

\begin{figure}
\includegraphics[angle=-90]{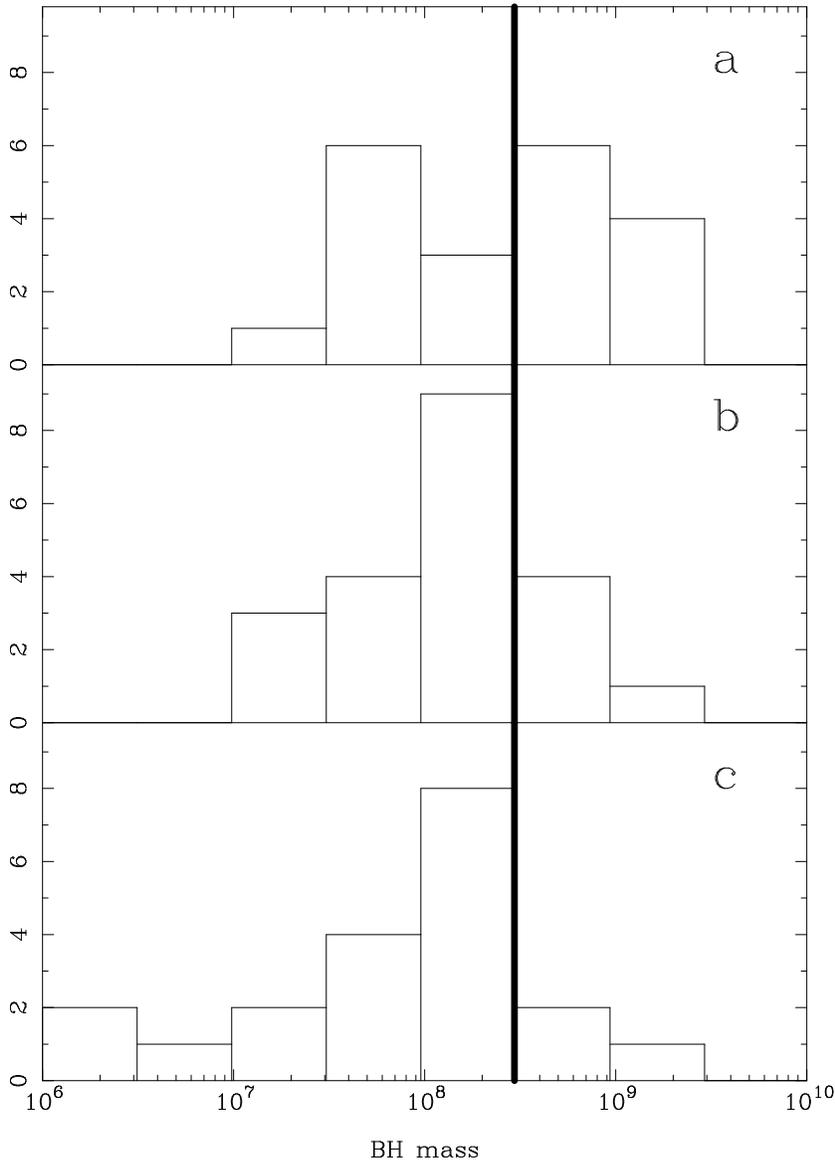}
\caption{The histogram for BH masses of the 20 quasars that have
quasi-simultaneous optical and near IR photometry. The four quasars
that have SEDs that indicate dust reddening have been excluded. The
photometry enables us to do a
proper K-correction for the flux at 5100\AA. The solid line represents
the lower limit on the mass of BHs in radio-loud quasars suggested
by \citet{lao00}. (a) shows
the 20 quasars using the B$_j$ magnitude to calculate
the flux at 5100\AA. (b) uses the simultaneous
photometry to calculate the K-corrected flux at 5100\AA.
(c) uses fluxes corrected to take into account the
fraction of the synchrotron that extends into the optical continuum
calculated by \citet{whi01}.}
\label{all_hist}
\end{figure}

\begin{enumerate}

\item All the sources in this sample are flat spectrum.  One might speculate
that the radio emission in particular is beamed, increasing the radio
flux relative to the optical.  The effect of the beaming would be that
the points in Figure~\ref{bh_R} are at higher $\cal R$ values than their
non-beamed values, but this should be
independent of BH mass unless the latter also effects the Lorentz factor
of the jet.  Interestingly, in a recent preprint, \citet{gu01} have shown
using a similar sample to the PHFS, that steep and flat spectrum
quasars have similar distributions of BH mass compared to radio
luminosity and radio loudness $\cal R$. 

\item The spectral energy distributions (SEDs) for radio-loud quasars
  vary greatly, making a uniform K-correction invalid. This means that
  for quasars at
different redshifts we are sampling different parts of the spectrum by
using the $B_j$ magnitude.

To investigate this effect, we can use
24 of the quasars with measured velocity profiles which also have
quasi-simultaneous photometry in seven bands covering the optical and
near IR \citep{fra00}. Four of these have evidence for dust reddening
which is discussed further in item~4. The original BH mass distribution of
these 20 unreddened quasars is shown in Figure~\ref{all_hist}a. Using the
simultaneous photometry we can now linearly
interpolate between these SED data points to obtain an accurate flux at
a rest wavelength of 5100\AA. The histogram of the results is shown in
Figure~\ref{all_hist}b. The range in BH masses has decreased, and
the lower limit proposed by \citet{lao00} is even more strongly violated.

\item It has been shown that the
synchrotron radiation from the jet can extend into the near IR and
optical \citep{whi01}. This will cause an increase in the observed
continuum flux but, since the jet emission is
beamed, would not contribute to the
ionizing flux seen by the BLR. \citet{whi01}
have fitted models to the quasi-simultaneous photometry that consist
of both a synchrotron component and a blue powerlaw (representing the
ionizing flux from the accretion disk). From these fits we can
calculate the fraction of the total emission that comes from the
accretion disk. The accretion disk flux will be less than the observed
flux if there is a significant synchrotron contribution present in the
optical. The results of this correction are shown
in Figure \ref{all_hist}c where some of the objects
with a large synchrotron component now have much lower BH masses.
Nearly all the quasars are below $10^9$M$_\odot$.

\item Dust may affect the observed luminosity of the   
  quasar. Dust will tend to redden the spectrum of the quasar and
  reduce the observed luminosity. The dust extinction is modeled as an
  exponential function of wavelength with flux at the blue wavelengths
  depleted significantly more than the red. For quasars where we have
  simultaneous photometry,  evidence of dust is shown by a steep
  decrease in flux at higher energies. 
  For the subset of quasars that have photometry we
  see evidence of dust in the SEDs of four of
  the quasars. We have not used these quasars in the previous
  analysis as their flux will be underestimated. After removing these,
  we still see the very
  large range in BH masses including low mass BHs.

\item The H$\beta$ line might be emitted from a disk-like structure,
with its line width reflecting the rotation of that disk
\citep{wil86}. Then, since it is
expected that flat spectrum sources are viewed face-on, the 
width of the H$\beta$ line may be underestimated.
In the \citet{gu01} sample the median and mean values for their flat
spectrum sample are lower than the steep spectrum sample.
This data partially supports the flattened disk
scenario (see Table~\ref{tab2}). However it cannot be the whole
story as the widths of the
emission lines of the flat spectrum quasars are still a significant fraction
of the steep spectrum ones and, the flat-spectrum sources need only to
be oriented at an average angle of 53$^{\circ}$ to the
line-of-sight to reconcile to difference in the median values of FWHM.
If we use the \citet{gu01} difference, this introduces a mean
difference of a factor of 2 in the BH mass.

\item Contamination of the AGN by the host galaxy light will effect
  all AGN to some extent, however it is obviously more significant for the
  objects where the host galaxy is resolved optically. The flux measured is the
  integrated flux over the entire galaxy. The central source flux
  which is required to calculate the BH mass, is much
  smaller. This effect will be present for all objects, but where the
  galaxy is resolved it will be the most pronounced as the host galaxy is a
  bigger percentage of the total observed flux. The result
  will be that the actual BH mass, for the AGN, is quite a bit smaller than
  calculated, so these masses are upper limits. The FWHM of the
  emission lines may also be effected by the host galaxy if it has
  strong H$\alpha$ or H$\beta$ emission lines. As we haven't resolved
  the broad and narrow components of the emission, the FWHM may be
  under-estimated due to a higher peak coming from the host galaxy
  emission. This is more likely to be a factor in the
  H$\alpha$ emission lines. The resolution of the spectra was not
  sufficient to consider contamination by weaker nearby emission lines
  such as \ion{Fe}{2} and \ion{N}{2}. It is noted that the FWHM
  velocity of the H$\beta$ line, after deconvolution, would need to
  increase by a factor of 
  10 to be consistent with the \citet{lao00} finding - an unlikely result.

\end{enumerate}

\subsection{Optical Properties of Quasars}

The results suggest that BH mass is not
connected to the radio-loudness of the objects. The question then
arises: Why have studies in the past produced a different result? 
Figure~\ref{bh_opt} shows that the range in values that we calculate
for the BH mass is largely a consequence of using the optical
luminosity to estimate the radius of the BLR. The velocity width of
the lines has a much smaller effect on the BH mass distribution due to
its smaller dynamic range. Therefore, the reason that previous studies
using this method have missed the low BH mass radio-loud objects is
because they have missed the low luminosity radio-loud objects.
We
believe that optical selection is biased against selecting low optical
luminosity, radio-loud quasars as shown in the PG sample \citep{hoo96}.

The selection criterion for the low redshift PG quasars used by
\citet{lao00} are: 
(1) UV excess (U$-$B$ < -0.44$).
(2) morphological criteria where the objects must have a `dominant
star-like appearance'. They are also bright with a limiting magnitude
$\sim$16.6. Therefore we obtain a sample of bright, blue, point-like objects.
 The key difference in the PHFS and the sample used by \citet{lao00}
 is that the PHFS has a large range in optical luminosity which produces the
 large distribution in BH mass whereas the radio-loud quasars selected
 on UV excess tend to have the characteristic of being optically
 luminous which therefore gives large BH masses. From the PHFS
 it is evident that not all radio-loud quasars are highly optically luminous.

\begin{figure}
\includegraphics[angle=-90]{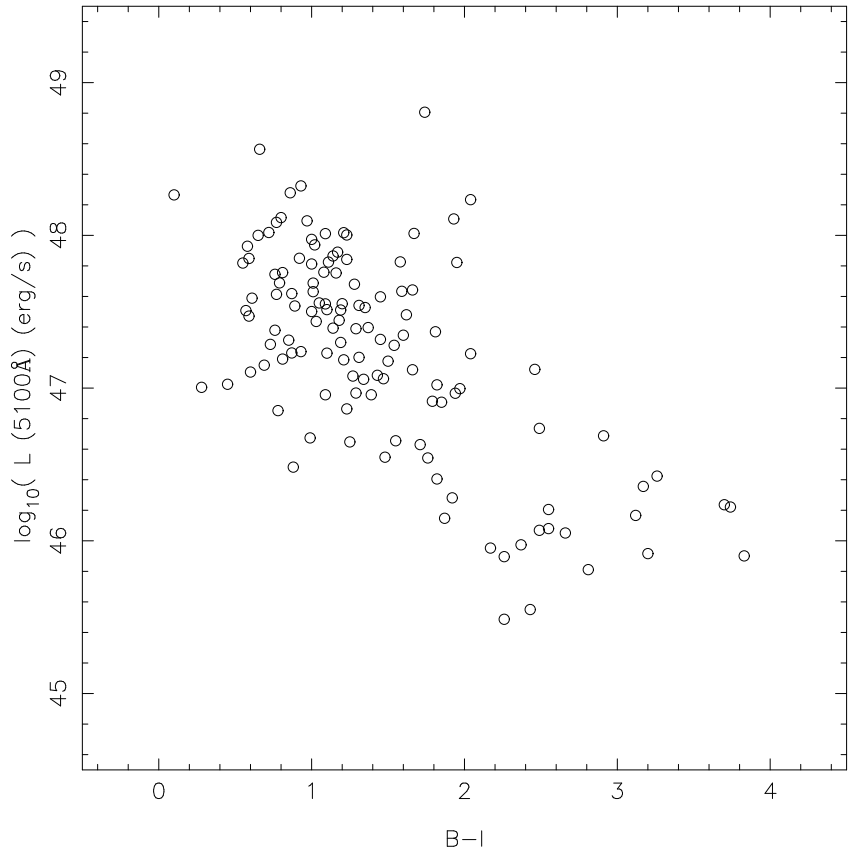}
\caption{Colour vs Optical Luminosity plot for the whole PHFS with
  quasi-simultaneous photometry \citep{fra00}. It shows that the
  redder objects are the ones with lower optical luminosity and would
  probably be missed in samples that have been selected based on
  optical colours or UV excess.}
\label{L_colour}
\end{figure}

Figure~\ref{L_colour} provides evidence of an additional correlation
which might effect selection into quasar samples. In
figure~\ref{L_colour}, which plots optical luminosity against colour
(B-I), low luminosity quasars are shown to be redder. Thus any
additional selection criterion based on colour will further bias
against low luminosity quasars.
The photometry of these red quasars indicates that there are three
types of reddening taking place. We find that a
proportion of these are red due to synchrotron extending into the
optical part of the spectrum and therefore boosting the red end of the
continuum. This has been discussed and corrected for in
section~\ref{bias}. This is an effect that will only be evident in radio-loud
quasars with high energy synchrotron jets and will not be observed in
radio-quiet quasars. Another subset of the redder objects in
the PHFS
have a substantial fraction of galactic light included in the
photometry so the 4000\AA\ break contributes to the red colour. This
occurs mainly in the low redshift 
population which is a small subsample of our estimated BH
masses. Most of the sources classified as `g' or `G' in Figure~\ref{bh_R}
do not have a significant galactic contribution in their
spectra (for example, a significant 4000\AA\ break),
and are obviously AGN due to their strong radio loudness and broad
emission lines. A
smaller proportion of the sample are very faint and show some evidence of dust
reddening where the blue end of the continuum is strongly suppressed.
In this case the BH masses will be underestimated but the previous two
effects lead to overestimation of the BH mass. Using the
simultaneous photometry we have identified 4 out of 24 dust-reddened objects in
the sample and have excluded them in the analysis shown in
Figure~\ref{all_hist}: we still get a population of quasars with
relatively low BH mass.

Why do red quasars have low optical luminosity, and therefore (under
our assumptions), low BH mass? We have argued that a significant
fraction of red quasars have synchrotron turning over at optical
wavelengths \citep{whi01}. This seems to be correlated with lower optical
emission which could be interpreted in two ways. Either, the
synchrotron emission remains unchanged but appears dominant only when
the blue disk emission is low, or secondly, when the blue disk
emission is low,
relatively more energy goes into the jet and the synchrotron is
increased. We do not have a large enough sample to establish the
correlation between low luminosity and redness unequivocally, but if
the correlation could be substantiated it might provide physically
interesting constraints on the jet formation mechanism.

Theoretical calculations considered by \citet{mei99} demonstrate
explicitly that it is not necessary to have a relatively massive BH to
produce powerful radio jets. In his model, based on current
accretion and jet-production theory, the strength of the radio emission
is tied to the rotation rate of the central BH. Therefore, it is quite possible
to have highly powered radio jets with small BH masses if the BH is
spinning rapidly.

\section{Conclusion}

The virial method for estimating BH mass in
conjunction with the empirical relation
between luminosity and radius \citep{kas00} has been applied to the
PHFS sample of flat spectrum radio sources. This is the first time the
analysis has been applied to an optically complete radio-selected
sample. The analysis is complimentary to the \citet{ho01} analysis
which concentrates on optically luminous sources used in longterm
monitoring programs. The errors and
applicability of this method have been discussed. The main result are
summarized as follows:
\begin{enumerate}
\item We find no evidence for a lower cut off in the masses of
  radio-loud quasars as previously suggested in the literature.
\item The previous lower boundary on mass of radio-loud quasars was an optical
selection effect. Low luminosity radio-loud AGN 
have been shown to have redder optical colours. These red quasars will
be selected against in optical surveys. 
\item We do not find any evidence for a relationship between radio
  properties and the BH mass of quasars. Further investigation of this
  relationship is hampered by unknown orientation effects.
\end{enumerate}

To obtain more accurate BH estimates and improve our understanding
of the relationship between the mass of the central BH and other
properties of quasars, we need large samples of BH masses measured
using the reverberation technique. This is now observationally
possible using multi-fibre spectrographs.

We wish to acknowledge the anonymous referee for careful reading of
the manuscript and valuable comments which lead to the improvement of
this paper.

\clearpage

\clearpage

\clearpage

\clearpage

\clearpage

\clearpage

\clearpage

\clearpage

\begin{table}
\label{tab1}
\caption{Observational Data for PHFS Quasars and Estimated BH Mass}
\begin{footnotesize}
\begin{tabular}{p{2.3cm}p{0.8cm}p{0.8cm}p{1.6cm}p{0.7cm}p{1.6cm}p{1.4cm}p{1.1cm}p{1.1cm}p{1.5cm}}
\hline
Name& z & $B_j$ &$\lambda L_{\lambda}$ (ergs~s$^{-1}$) & 5GHz
flux (Jy) & P$_{rad}$ (5GHz) (W/Hz)&  $\cal R$ &H$\beta$ FWHM ($km s^{-1}$) & H$\alpha$ FWHM ($km s^{-1}$)& BH mass
(M$_\odot$) \\
\hline\hline
PKS 0114$+$074 & 0.343 & 22.14 & 1.05$\times 10^{43}$ & 0.67 &
1.74$\times 10^{26}$ & 1.11$\times 10^5$ & 2515 &... & 6.30$\times 10^6$ \\ 
PKS 0153$-$410 & 0.226 & 19.41& 5.41$\times 10^{43}$ & 0.94 &
1.02$\times 10^{26}$ & 1.26$\times 10^4$ & 3385 & 1932& 3.59$\times 10^7$ \\ 
PKS 0221$+$067 & 0.510 & 20.76 & 8.75$\times 10^{43}$& 0.77 &
4.66$\times 10^{26}$ & 3.59$\times 10^4$ & 2118 &... &1.97$\times 10^7$ \\ 
PKS 0327$-$241 & 0.888 & 19.39 & 1.03$\times 10^{45}$ & 0.73 &
1.48$\times 10^{27}$ & 9.63$\times 10^3$ & 4007 &... &3.96$\times 10^8$ \\ 
PKS 0454$+$066 & 0.405 & 19.79 & 1.30$\times 10^{44}$& 0.44 &
1.63$\times 10^{26}$ & 8.39$\times 10^3$ &... &2128 &3.77$\times 10^7$ \\ 
PKS 0502$+$049 & 0.954 & 18.7 & 2.28$\times 10^{45}$& 0.82 &
1.94$\times 10^{27}$ & 5.73$\times 10^3$ & 4184 &... &7.53$\times 10^8$\\ 
PKS 0912$+$029 & 0.427 & 19.56 & 1.80$\times 10^{44}$ & 0.46 &
1.90$\times 10^{26}$ & 7.10$\times 10^3$ & 2678 & 3152 &5.23$\times 10^7$ \\ 
PKS 0921$-$213 & 0.052 & 16.4 & 4.25$\times 10^{43}$& 0.42 &
2.23$\times 10^{24}$ & 3.53$\times 10^2$ &... &7238 &1.99$\times 10^8$ \\ 
PKS 0925$-$203 & 0.348 & 16.35 & 2.25$\times 10^{45}$ & 0.70 &
1.87$\times 10^{26}$ & 5.61$\times 10^2$ & 2611 & 2217 &2.90$\times 10^8$ \\ 
PKS 1016$-$311 & 0.794 & 17.58 & 4.28$\times 10^{45}$ & 0.65 &
1.03$\times 10^{27}$ & 1.62$\times 10^3$ & 3416 &... & 7.79$\times 10^8$ \\ 
PKS 1020$-$103 & 0.196 & 15.07 & 2.19$\times 10^{45}$ & 0.49 &
3.93$\times 10^{25}$ & 1.21$\times 10^2$ & 7920 & 6138 & 2.62$\times 10^9$ \\ 
PKS 1034$-$293 & 0.312 & 15.94 & 2.60$\times 10^{45}$ & 1.51 &
3.21$\times 10^{26}$ & 8.30$\times 10^2$ & 3463 &... & 5.65$\times 10^8$ \\ 
PKS 1036$-$154 & 0.525 & 21.8 & 3.57$\times 10^{43}$ & 0.78 &
5.03$\times 10^{26}$ & 9.47$\times 10^4$ & 5216 & 3671 & 6.38$\times 10^7$ \\ 
PKS 1101$-$325 & 0.355 & 16.45  & 2.14$\times 10^{45}$ & 0.73 &
2.04$\times 10^{26}$ & 6.42$\times 10^2$ & 3135 &2949 & 4.04$\times 10^8$ \\ 
PKS 1106$+$023 & 0.157 & 18.01& 9.22$\times 10^{43}$  & 0.50 &
2.54$\times 10^{25}$ & 1.85$\times 10^3$ &... &2632 &4.52$\times 10^7$ \\ 
PKS 1107$-$187 & 0.497 & 22.44& 1.76$\times 10^{43}$  & 0.50 &
2.86$\times 10^{26}$ & 1.09$\times 10^5$ &... &2361 &1.14$\times 10^7$ \\ 
PKS 1128$-$047 & 0.266 & 21.41 & 1.21$\times 10^{43}$& 0.90 &
1.37$\times 10^{26}$ & 7.63$\times 10^4$ &... &2186 &7.52$\times 10^6$ \\ 
PKS 1136$-$135 & 0.557 & 16.3 & 6.43$\times 10^{45}$ & 2.22 &
1.63$\times 10^{27}$ & 1.70$\times 10^3$ & 2275 &2731 &4.60$\times 10^8$ \\ 
PKS 1200$-$051 & 0.381 & 16.42 & 2.55$\times 10^{45}$ & 0.46 &
1.49$\times 10^{26}$  & 3.94$\times 10^2$ & 2361 &1803 &2.59$\times 10^8$ \\ 
PKS 1226$+$023 & 0.158 & 12.93 & 1.01$\times 10^{46}$ & 40.0&
2.05$\times 10^{27}$   & 1.37$\times 10^3$ & 4040 &... &1.98$\times 10^9$ \\ 
PKS 1237$-$101 & 0.751 & 17.46 & 4.23$\times 10^{45}$ & 1.13 &
1.58$\times 10^{27}$  & 2.52$\times 10^3$ & 5392 &... &1.93$\times 10^9$ \\ 
PKS 1254$-$333 & 0.190 & 17.05 & 3.31$\times 10^{44}$ & 0.54 &
4.06$\times 10^{25}$  & 8.25$\times 10^2$ & 7800 &... &6.78$\times 10^8$ \\ 
PKS 1302$-$102 & 0.286 & 15.71& 2.68$\times 10^{45}$ & 1.0 &
1.77$\times 10^{26}$  & 4.45$\times 10^2$ & 3417 &2439 &5.61$\times 10^8$ \\ 
PKS 1352$-$104 & 0.332 & 17.6 & 6.43$\times 10^{44}$ & 0.98 &
2.37$\times 10^{26}$  & 2.49$\times 10^3$ & 2814 &... &1.40$\times 10^8$ \\ 
PKS 1359$-$281 & 0.803 & 18.71 & 1.55$\times 10^{45}$& 0.67 &
1.09$\times 10^{27}$  & 4.72$\times 10^3$ & 1889\tablenotemark{*} &... &1.17$\times 10^8$ \\ 
PKS 1450$-$338 & 0.368 & 22.52 & 8.61$\times 10^{42}$ & 0.54 &
1.63$\times 10^{26}$ & 1.27$\times 10^5$ &... &1824 &4.14$\times 10^6$ \\ 
PKS 1509$+$022 & 0.219 & 19.83 & 3.44$\times 10^{43}$ & 0.54 &
5.46$\times 10^{25}$  & 1.07$\times 10^4$ & 6551 &... &9.80$\times 10^7$ \\ 
PKS 1510$-$089 & 0.362 & 16.21& 2.78$\times 10^{45}$ & 3.25 &
9.46$\times 10^{26}$  & 2.29$\times 10^3$ & 2796 &... &3.86$\times 10^8$ \\ 
PKS 1546$+$027 & 0.415 & 18.54 & 4.34$\times 10^{44}$ & 1.42 &
5.53$\times 10^{26}$  & 8.56$\times 10^3$ & 3893 &2831 &2.04$\times 10^8$ \\ 
PKS 1555$-$140 & 0.097 & 16.99 & 8.77$\times 10^{43}$ & 0.83 &
1.57$\times 10^{25}$  & 1.20$\times 10^3$ &... &2001 &2.53$\times 10^7$ \\ 
PKS 1706$+$006 & 0.449 & 22.8& 1.02$\times 10^{43}$ & 0.38 &
1.75$\times 10^{26}$  & 1.16$\times 10^5$ &... &2086 &6.08$\times 10^6$ \\ 
PKS 1725$+$044 & 0.296 & 18.2 & 2.90$\times 10^{44}$  & 1.21 &
2.30$\times 10^{26}$  & 5.33$\times 10^3$ & 2638 &2399 &7.07$\times 10^7$ \\ 
PKS 1954$-$388 & 0.626 & 17.82 & 2.04$\times 10^{45}$ & 2.0 &
1.89$\times 10^{27}$  & 6.21$\times 10^3$ & 3293 &... &4.32$\times 10^8$  \\ 
PKS 2004$-$447 & 0.240 & 18.09 & 2.07$\times 10^{44}$ & 0.65 &
7.96$\times 10^{25}$  & 2.59$\times 10^3$ & 1939 &1602 &3.01$\times 10^7$ \\ 
PKS 2059$+$034 & 1.012 & 17.64 & 6.90$\times 10^{45}$ & 0.75 &
2.02$\times 10^{27}$ & 1.97$\times 10^3$ & 3815 &... &1.36$\times 10^9$ \\ 
PKS 2120$+$099 & 0.932 & 20.16 & 5.65$\times 10^{44}$ & 0.5 &
1.13$\times 10^{27}$ & 1.34$\times 10^4$ & 3083 &... &1.54$\times 10^8$  \\ 
PKS 2128$-$123 & 0.499 & 15.97& 6.88$\times 10^{45}$  & 2.0 &
1.16$\times 10^{27}$  & 1.13$\times 10^3$ & 4652 &4220 &2.02$\times 10^9$ \\ 
PKS 2143$-$156 & 0.698 & 17.24 & 4.41$\times 10^{45}$ & 0.82 &
9.80$\times 10^{26}$  & 1.49$\times 10^3$ & 836 &... &4.78$\times 10^7$  \\ 
PKS 2329$-$415 & 0.671 & 18.2 & 1.67$\times 10^{45}$ & 0.47 &
5.15$\times 10^{26}$  & 2.07$\times 10^3$ & 4952 &... &8.49$\times 10^8$  \\ 

\hline
\end{tabular}
\tablenotetext{*}{H$\gamma$ FWHM, no data for H$\alpha$ and H$\beta$ available}
\end{footnotesize}
\end{table}

\clearpage

\begin{table}
\begin{center}
\caption{Average H$\beta$ FWHMs for flat spectrum and steep spectrum
  quasars from \citet{gu01}}
\label{tab2}
\begin{tabular}{p{3.3cm}p{2.9cm}p{2.9cm}}
\hline  & mean (km s$^{-1}$)  & median (km s$^{-1}$)\\
\hline
\hline Flat Spectrum & 3632 & 3600 \\
Steep Spectrum & 5240 & 4500  \\
\tableline
\end{tabular}
\end{center}
\end{table}
 
\end{document}